# Emerging Device Applications for Semiconducting Two-Dimensional Transition Metal Dichalcogenides

*Deep Jariwala,[†] Vinod K. Sangwan,[†] Lincoln J. Lauhon,[†] Tobin J. Marks,[†,‡] and Mark C. Hersam[†,‡,∥],\**

[†]Department of Materials Science and Engineering, Northwestern University, Evanston, Illinois 60208, U.S.A., [‡]Department of Chemistry, Northwestern University, Evanston, Illinois 60208, U.S.A., [∥]Department of Medicine, Northwestern University, Evanston, Illinois 60208, U.S.A.

*Address correspondence to: m-hersam@northwestern.edu

ABSTRACT: With advances in exfoliation and synthetic techniques, atomically thin films of semiconducting transition metal dichalcogenides have recently been isolated and characterized. Their two-dimensional structure, coupled with a direct band gap in the visible portion of the electromagnetic spectrum, suggests suitability for digital electronics and optoelectronics. Towards that end, several classes of high-performance devices have been reported along with significant progress in understanding their physical properties. Here, we present a review of the architecture, operating principles, and physics of electronic and optoelectronic devices based on ultrathin transition metal dichalcogenide semiconductors. By critically assessing and comparing the performance of these devices with competing technologies, the merits and shortcomings of this emerging class of electronic materials are identified, thereby providing a roadmap for future development.

VOCABULARY: **Mobility** - A term used to characterize the speed of electric charge carriers upon application of an electric field. **Field effect transistor** - A three-terminal electronic device that functions as a voltage-controlled switch. **Diode** - An electronic device with highly asymmetric current-voltage characteristics that predominantly allows current flow for only one bias polarity. **Van der Waals heterostructure** - Vertical stack of disparate two-dimensional materials interacting only via weak van der Waals forces. **Photodetector** – An optoelectronic device that produces electrical current in response to optical excitation.

The relentless miniaturization of silicon-based electronics coupled with the advent of graphene has led the electronic materials community towards atomically thin two-dimensional semiconductors.[1] While graphene has attracted significant attention, it lacks a band gap and thus is unsuitable for digital electronic applications.[1-5] Consequently, significant effort has been devoted to identifying alternative two-dimensional semiconductors. Several classes of non-graphene layered compounds have received recent attention including boron nitride, metal chalcogenides, oxides, hydroxides, and oxychlorides. A comprehensive list of all known layered Van der Waals solids has been reported in a number of recent reviews.[6-11] However, only a few of these layered materials can be classified as semiconductors, and even fewer have been successfully isolated as air-stable, high-quality, two-dimensional crystals.

Transition metal dichalcogenides (TMDCs) are among the most studied layered compounds that have been isolated in monolayer form. Compounds in the TMDC family exhibit a wide range of electrical properties, depending on polytype and the number of transition metal d-electrons, and include metallic,[12] half-metallic,[13] semiconducting,[14-16] superconducting,[17] and charge density wave[18] behavior. In particular, molybdenum and tungsten based TMDCs are semiconductors with band gaps ranging from the visible to the near-infrared. Besides Mo and W, chalcogenides of Ti, Sn, and Zr are also predicted to be semiconducting but little to no experimental evidence exists on their isolation in monolayer form, stability, or performance in devices.[19-23] Thus, Mo and W chalcogenides have been the most heavily investigated among the post-graphene two-dimensional materials.

This review will primarily focus on semiconducting TMDCs based on Mo and W and their applications in devices. First, the unique physical properties of semiconducting TMDCs that are most relevant in device applications will be highlighted such as crystal structure, symmetry, the thickness-dependent evolution of electronic and phonon structure, as well as the effects of quantum confinement. A series of emerging electronic and optoelectronic device applications will then be delineated, including field-effect transistors (FETs), heterostructure junctions, photodetectors, photovoltaics, and sensors. In order to provide context and identify the most productive opportunities for future research, TMDC device metrics will be compared and contrasted with those of competing semiconductors and pre-existing technologies.[3, 24-33]

**Physical properties**

TMDCs have the chemical formula $MX_2$, where M is a transition element and X is a chalcogen.[12, 34, 35] Various permutations of these elements result in over 40 different compounds, although not all of them are layered solids.[6, 8, 12, 36] A monolayer is defined here as a hexagonally ordered plane of metal atoms sandwiched between two hexagonally ordered planes of chalcogen atoms, where the formal oxidation states of the metal and chalcogen elements are +4 and -2, respectively.[35] The structure of layered TMDCs is similar to that of graphite and each layer has a thickness of 6 – 7 Å with strong in-plane covalent bonding and weak out-of-plane van der Waals interactions (Figure 1a). Bulk TMDCs are found in several structural polytypes depending on the stacking order of the layers,[35] while single layers of TMDCs are found in two polytypes, depending on the position of the chalcogen with respect to the metal element in the X-M-X structure. The TMDC monolayer polytypes can be either trigonal prismatic ($D_{3h}$ point group, honeycomb motif) or octahedral ($D_{3d}$ point group, centered-honeycomb motif) also referred to as 2H (1H for a monolayer) and 1T, respectively (Figure 1b).[8, 35, 37] The half-filled *d*-orbitals of group VIB transition metals (*i.e.*, M = Cr, Mo, W) results in semiconducting behavior with decreasing band gap as the chalcogen atomic number is increased (*i.e.*, X = S, Se, Te). These semiconducting TMDCs are found primarily as the 2H polytype, although reversible transitions to the 1T polytype have been achieved via chemical processing.[38]

Weak van der Waals interactions between $MX_2$ layers facilitate isolation of single layers via mechanical[39-41] or chemical exfoliation.[6, 22, 23, 42, 43] For example, Figure 1c shows an optical image of single layer (SL) $MoS_2$ exfoliated *via* the "Scotch-tape" method. The electrical properties of semiconducting TMDCs also depend on the number of layers due to quantum confinement effects and changes in symmetry.[44, 45] Several groups have predicted[46-49] and verified[38, 44, 45, 50] the transition from an indirect band gap at the Γ-point to a direct band gap at the K-point of the Brillouin zone as the semiconducting TMDC thickness is reduced to a single layer. In particular, bulk $MoS_2$ has an indirect band gap of 1.2 eV, whereas single layer $MoS_2$ has a direct band gap of 1.9 eV (Figure 1d,e)[46] that results in enhanced photoluminescence (PL) by up to 4 orders of magnitude.[44] Similar evidence for a direct band gap transition has also been reported for single layer $MoSe_2$, $WS_2$, and $WSe_2$.[51, 52]

Recent theoretical work[49, 53, 54] and experiments[55-57] also suggest that optical transitions in ultrathin TMDC samples are dominated by excitons rather than direct interband transitions. For example, the optical absorption spectra of both single-layer and bilayer $MoS_2$ exhibit two distinct peaks that have been assigned to A and B excitons.[44] The excition binding energy is predicted to be as high as 0.5-0.9 eV for a monolayer and ~0.4 eV for a bilayer.[49, 53, 54] The exciton binding energy is dependent on the strain and surrounding dielectric environment, which can be detected as shifts in the emission peak.[58-60] Furthermore, gate-dependent photoluminescence measurements on single-layer semiconducting TMDCs suggest that an exciton can be charged with an extra electron or hole to form trions (Figure 1f, g).[55, 56] In addition, the unique symmetry of semiconducting TMDCs suggests promise for using a new state variable (*i.e.*, valley quantum numbers) for logic operations in so-called valleytronic devices.[61] Earlier, the valley degree of freedom had only been accessed by applying stain[62] or magnetic fields[63] in non-TMDC materials, whereas, as discussed below, the properties of TMDCs allow access to valley degrees of freedom *via* circularly polarized light. In contrast to graphene, single-layer TMDCs have a honeycomb structure with M and $X_2$ atoms at alternating corners, so the degeneracy of the K and K' points is lifted.[35, 61, 64] Furthermore, time reversal symmetry requires that split valleys have charge carriers with opposite spin. The net result is coupling between the spin and valley degrees of freedom, allowing confinement of charge carriers in a particular valley (Figure 1h).[61, 64] Recently, multiple groups have demonstrated selective valley population by exciting edge electrons with circularly polarized light in monolayers of $MoS_2$ and $WSe_2$.[64-68]

The phonon structure of semiconducting TMDCs is also sensitive to the number of layers and strain as well as temperature and carrier concentration. The Raman spectra of TMDCs contain two predominant and distinct peaks : (1) The out-of-plane $A_{1g}$ mode where the top and bottom X atoms are moving out of plane in opposite directions while M is stationary; (2) The in-plane degenerate $E^1_{2g}$ mode where the M and X atoms are moving in plane in opposite directions.[69] The addition of extra layers leads to stiffening of the out-of-plane phonon modes and relaxation of in-plane bonding, resulting in a blue shift of the $A_{1g}$ mode and a red shift of the $E^1_{2g}$ mode.[69, 70] Both of these modes undergo a red shift as well as spectral broadening with increasing temperatures that is attributed to anharmonic contributions to the interatomic potentials.[71-73] The $A_{1g}$ mode is also sensitive to carrier density and undergoes a red shift as well as peak broadening with electron doping while the $E^1_{2g}$ mode remains insensitive to doping, suggesting stronger electron-phonon coupling with the out-of-plane vibrational mode.[74] The same effect is also evident in resonant Raman spectra of TMDCs where the intensity of the $A_{1g}$ mode is greatly enhanced under resonance conditions while the enhancement vanishes with reduced thickness due to reduced electron-$A_{1g}$ coupling in thinner samples.[70] Resonance conditions have also led to the observation of a new second-order longitudinal acoustic (LA) phonon mode, the intensity of which increases upon reduction in the sample thickness.[70, 75] While the $E^1_{2g}$ mode remains insensitive to doping, it is very sensitive to applied strain. Applying uniaxial strain lifts the degeneracy of the $E^1_{2g}$ mode, leading to red shifting and splitting into two distinct peaks for strain of just 1%.[59] The unique electronic band structure and lattice vibration properties of TMDCs suggest new opportunities for electronic, optoelectronic, and sensing devices as will be outlined below.

**<u>Digital electronic devices</u>**

Digital electronics requires the reliable creation, storage, and reading of distinct voltage states. In Boolean logic, two states are required, corresponding to true '1' or false '0'. Digital electronic circuits are comprised of large assemblies of logic gates, which are electronic implementations of the Boolean logic functions. Each logic gate typically consists of several transistors, making the transistor the fundamental component of modern digital electronics. Most digital electronic circuits use the metal-oxide-semiconductor field-effect transistor (MOSFET) as the basic component. The two discrete voltage states described above can be achieved by having two distinct states of conductance in a MOSFET separated by multiple orders of magnitude. This high on/off current ratio requirement is most easily achieved at room temperature in semiconducting materials with band gaps greater than ~1 eV. The substantial band gaps of semiconducting TMDCs suggest that high on/off current ratios can be achieved, unlike in gapless graphene, and thus digital electronics based on 2D materials is possible. This section will first focus on the development of ultrathin semiconducting TMDC-based FETs, followed by a review of advanced digital circuit elements such as inverters and logic gates.

**Field-effect transistors**

Although the electrical properties of bulk semiconducting TMDCs have been investigated for decades,[15, 76] they were first used in FETs only over the past ~10 years.[77] It was only after the landmark paper on graphene by the Manchester group[78] that monolayers of semiconducting TMDCs were investigated in FETs.[39] In early work, the field-effect mobility for single layer (SL) MoS$_2$ was found to be at least 3 orders of magnitude lower than graphene.[39] The interest in monolayer semiconducting TMDC FETs was renewed in 2011 when top-gated SL-MoS$_2$ FETs (Figure 2a inset) with moderate mobilities (~60-70 cm$^2$/Vs[79-81]), large on/off ratios (~10$^8$), and low sub-threshold swings (74 mV/dec) were demonstrated at room temperature (Figure 2a).[16] The two-dimensional channel coupled with a large band gap (> 1 eV) and ultrathin top gate dielectric allowed superior gate control, thus enabling small off currents and large switching ratios. This work also suggested that the presence of the top dielectric enhanced the carrier mobility.

Since charge carriers in ultrathin TMDCs are weakly screened, the carrier mobility is sensitive to the interfaces on both sides of the semiconductor film. This effect is particularly evident in vacuum *versus* ambient measurements[82-84] of unencapsulated SL-MoS$_2$ FETs (Figure 2b). A significant negative shift in threshold voltage and higher on-currents are observed under vacuum conditions, indicating p-type doping and charge scattering caused by atmospheric adsorbates.[82] This performance degradation due to adsorbate interaction is reversible in the case of TMDCs by reinserting the devices into vacuum conditions unlike in organic semiconductors where this degradation is irreversible. Similarly, large mobility differences have been observed in bare *versus* top dielectric encapsulated devices,[16, 80, 85-87] bottom-gated devices on different dielectrics,[88-92] and pre-bias *versus* post-bias stressing.[93] In addition, electrical double layer gating using ion gels lead to ambipolar transport[85, 94, 95] in otherwise n-type $MoS_2$ and $WS_2$ FETs, and increased carrier mobilities *via* increased screening of charged impurities. The ambipolar behavior has been attributed to the ultrahigh capacitance of the electrical double layer, which allows tuning of the Fermi level across the band gap to allow carrier injection into both the valence band and conduction band (Figure 2c).[94] Mobility increases have also been observed with increasing TMDC thickness due to screening of charged substrate impurities in addition to accessing the third dimension for charge transport.[88, 96-99] Therefore, to date, few-layer samples have exhibited the highest room temperature mobility values (>100 $cm^2$/Vs) in TMDC FETs.[88, 100]

The sensitivity of carrier mobilities to the local dielectric environment has inspired more thorough investigations of their role in charge transport. The first reports of temperature-dependent transport in ultrathin $MoS_2$ revealed variable range hopping[98] or thermally activated[101] transport as opposed to the band-like transport observed in bulk single crystals.[15] Following improvements in sample quality and device processing, band-like transport was observed in SL-$MoS_2$ with phonon scattering dominant for T > 100 K and charged impurity scattering dominant for T < 100 K.[82,80, 102] Similar behavior has also been reported on few-layer $MoS_2$[103] and $MoSe_2$.[104] A transition from variable range hopping to band-like transport with increasing carrier density was also observed in SL-$MoS_2$.[80, 102] In this case, the mobility ($\mu$) follows an inverse power law relation with temperature ($\mu \alpha T^{-\gamma}$) for T > 100 K, where the exponent $\gamma$ decreases from 1.4 to 0.7 after encapsulation with $HfO_2$. In addition, the value of $\gamma$ also falls from 1.7 in monolayer (Figure 3a) to 1.1 in bilayer unencapsulated devices.[102] In both of these examples, the presence of a top layer quenches/stiffens the homopolar $A^1_g$ mode to reduce $\gamma$ as predicted by theory (Figure 3 b).[105] However, the experimentally observed reduction in γ is much larger than the predicted value, which suggests that additional scattering mechanisms are active at higher temperatures.[106] Most theoretical work[105-108] has accounted only for intrinsic $MoS_2$ phonons, leading to the conclusion that longitudinal acoustic phonon scattering is dominant. More recent studies have considered the influence of remote interfacial phonons from the oxide dielectric[109] (akin to graphene[110]), which has led to better agreement with experimental observations.

While it is clear that phonon scattering is important at high temperatures, the disagreement between the theoretically predicted intrinsic mobility and the experimentally measured mobility in the best performing devices warrants further discussion. Theoretical studies have now predicted that the intrinsic mobility at room temperature in SL-$MoS_2$ is ~300-400 $cm^2$/Vs at high carrier densities of ~$10^{13}$/$cm^2$.[105, 107] This value is approximately 5 times greater than the largest experimentally reported values at 300 K,[82] suggesting that room temperature mobilities are further degraded by charged impurities and sub-optimal sample quality.[111] While early calculations neglected impurities/defects and thus overestimated mobility values, the qualitative dependence of mobility on carrier density does fall in line with experimental observations. In particular, the mobility has a weak dependence on carrier density at all temperatures except T < 10 K, where the dependence is strong.[108] This temperature dependence can be explained by charged impurity dominated scattering at low temperatures, where a higher carrier density better screens the charges and leads to higher mobility. It should be noted that all of the above studies focused on electronic transport in the low field limit. Although initial theoretical work on field-dependent mobility[107, 109] and high-field transport experiments[112-114] appear promising, a detailed study on the field dependence of mobility in ultrathin chalcogenides is currently absent.

Flicker noise or 1/f noise can be a limiting factor for electronic device applications. In particular, the all-surface structure of ultrathin TMDCs makes them sensitive to random perturbations in the environment. In a recent study, the lowest Hooge parameter in high mobility SL-MoS$_2$ was found to be 0.005 in vacuum, which is comparable to the noise amplitudes of carbon nanotubes and graphene.[83] However, since the Hooge parameter increases by as much as 3 orders of magnitude in ambient conditions, noise studies confirm the sensitivity of SL-MoS$_2$ to adsorbates and trapped charges, further suggesting that encapsulation strategies[115] may be effective in reducing the noise amplitude in TMDCs.

The contact resistance from metal electrodes plays an important role in determining the conductance of fully fabricated devices fabricated from reduced dimension materials. In the case of ultrathin MoS$_2$, commonly used Au contacts show linear current-voltage curves, suggesting ohmic contacts to MoS$_2$[16, 82, 98, 116] (Figure 4a). This result is counter-intuitive since Au has a high work function (5.1 eV) and thus should form a Schottky barrier with n-type MoS$_2$ (electron affinity ~ 4.0 eV). However, the width of the resulting Schottky barrier is exceptionally narrow due to the atomically thin TMDC, which enables low resistance tunneling through the barrier.[99] The net result is linear or electrically ohmic transport for ultrathin MoS$_2$ FETs even with high work function metal contacts.[99] Systematic studies of various metal contacts in ultrathin semiconducting TMDC transistors have also been performed both theoretically and experimentally.[99, 117-122] These studies found that low work function metals (*e.g.*, titanium and scandium) form lower resistance contacts than do high work function metals, leading to higher drain currents (Figure 4b).[99] The use of an ultrathin MgO[123] or TiO$_2$[124] barrier between a ferromagnetic metal and MoS$_2$ also minimizes Schottky barrier formation. Similarly, contact doping with nitrogen dioxide (NO$_2$),[125] elemental potassium,[126] and polyethylenimine (PEI)[127] has also been achieved in semiconducting TMDC FETs.

Nevertheless, even with an optimal contacting scheme, the contact resistance ultimately dominates charge transport with reduced channel dimensions.[128] This effect manifests itself as reductions in drain current and mobility at channel lengths below 500 nm,[128, 129] although quantum transport simulations predict excellent length scaling down to 10-15 nm.[130-132] The contact resistance of Au and Ni was also found to be a strong function of the applied gate bias,[102, 129] which suggests the presence of gate-tunable Schottky barriers (Figure 4c). It is therefore likely that field-effect mobilities reported in 2-probe devices are in general underestimated due to these contact effects. The reported values of contact resistance with Au[102, 129, 133] also vary widely, suggesting that changes in the surface cleanliness, sample quality, annealing, vacuum level, and method of evaporation can also have significant effects on the metal-semiconductor contact.

Overall, this section highlights several aspects of ultrathin semiconducting TMDC FETs that require deeper understanding and optimization to compete with conventional materials (*e.g.*, Si and GaAs) in high-performance transistors. However, in the near term, semiconducting TMDCs show promise in flexible, stretchable, and/or transparent electronics. Figure 5 shows a comparison of field-effect mobilities and on/off ratios for all candidate semiconductors that are being evaluated for these unconventional macroelectronic applications. As is evident from this figure, despite a significantly shorter history, semiconducting TMDCs have comparable mobilities and on/off ratios to organics,[26, 27, 30] amorphous oxide semiconductors,[134, 135] and semiconducting carbon nanotube networks.[136, 137] The availability of flexible contacts and dielectrics (*e.g.*, graphene, h-BN, and ion gels) have also enabled the fabrication of flexible, ultrathin electronic devices entirely from layered two-dimensional materials.[89] These initial studies demonstrate that ultrathin semiconducting TMDC FETs on flexible substrates are robust to mechanical bending with the important device parameters (*e.g.*, mobility, on/off ratio, and on-current) remaining within a desirable range.[138, 139] Flexible FETs from chemical vapor deposition (CVD) grown $MoS_2$ have also been demonstrated using ion gel dielectrics, which suggests large-area scalability of this device class.[140]

**Inverters and logic gates**

Since semiconducting TMDCs show promise as high performance FETs, researchers have begun integrating them into more complex, functional digital circuitry including inverters and logic gates. An inverter typically consists of two FETs in series, and is designed to convert logical 0 (low input voltage) to logical 1 (high output voltage) and *vice versa*. The common electrode between the two FETs serves as the output, while the gates of the FETs serve as the input (Figure 6a inset). An important metric for an inverter is the voltage gain, which is characterized by the negative derivative of the output plot (Figure 6a).[141] Single-layer $MoS_2$ inverters were first reported in 2011.[141] Thereafter, several reports of inverters based on a variety of semiconducting TMDCs have been published with gains varying between 2 and 16.[21, 142-144] A gain greater than unity is required in circuits consisting of cascaded inverters such as ring oscillators. In particular, a ring oscillator possesses an odd number of inverters in series to produce high frequency clock signals (Figure 6b).[142] Ring oscillators have been successfully demonstrated with $MoS_2$, leading to operating frequencies up to 1.6 MHz with stage delays of ~60 ns (Figure 6c).[142] Although this performance is inferior to ring oscillators based on conventional silicon, graphene,[145] or carbon nanotubes,[146] it is superior to current generation organics and amorphous oxide semiconductors,[147-149] which strengthens the case for semiconducting TMDCs in unconventional electronic applications. Similarly, NAND gates and static random access memory (SRAM) cells containing 3 and 4 FETs, respectively, have been demonstrated with ultrathin $MoS_2$.[142, 143]

A majority of the above reports are based on mechanically exfoliated TMDC flakes. However, practical manufacturing will require the production of large-area materials *via* scalable approaches. Towards that end, several groups have demonstrated the synthesis of atomic layers and multilayers of TMDCs using CVD techniques,[150-155] thus enabling fabrication of large-area electronics from atomically thin TMDC semiconductors (Figure 6d).[143] Note, however, that the carrier mobility in the CVD grown materials is about an order of magnitude lower than in mechanically exfoliated flakes and thus significant improvements in growth techniques will be required.[152, 153, 156] Despite significant progress in a short time span, all TMDC-based devices and circuits are unipolar in character thus far, which leads to high power consumption in inverters and logic gates. In contrast, complementary metal-oxide-semiconductor (CMOS) circuits enable low static power consumption, logic gate cascading, and highly integrated circuitry. Therefore, CMOS-based circuits using semiconducting TMDCs represent a key future goal for this field. In addition to controllable complementary doping schemes (*i.e.*, both n-type and p-type), methods for tailoring threshold voltage for complementary enhancement-mode FETs will be required.[157] Controlling doping and threshold voltages in atomically thin semiconductor devices has presented a major challenge since even a fractional percentage of substitutional doping can significantly degrade carrier mobility or alter the band structure. On the other hand, charge transfer doping by non-covalent functionalization/adsorption of dopant species has been effective in preserving the material quality, but this approach presents issues of long term stability. While the above challenges have plagued the field of carbon nanomaterials, naturally occurring chalcogen vacancies in TMDCs generate free carriers and, thus, controlled substitution/vacancy formation (without altering the band structure) may present a viable path forward in addition to stable, non-covalent charge transfer doping schemes. The reduced dimensionality of the conduction channel presents a similar challenge with threshold voltage control since the addition of more charges degrades carrier mobility. A combined approach involving doping as well as work function control of the gate electrode may present a viable solution for engineering threshold voltages in semiconducting TMDC field-effect devices.

**Junctions and heterostructures**

In addition to devices and circuits based on individual semiconducting TMDC materials, these materials present opportunities for unique device geometries based on junctions and heterostructures.[158, 159] Such heterojunction devices have been widely developed for crystalline III-V semiconductors for high frequency applications. These III-V semiconductor heterostructures have been historically grown by epitaxial crystal growth methods (*e.g.*, molecular beam epitaxy), which limits the range of possible heterojunctions to those materials that possess closely matched lattice parameters. On the other hand, the relatively weak van der Waals bonding between TMDCs relaxes this constraint and thus suggests a broader range of possible junctions and heterostructures.[10]

The use of ultrathin materials in vertical heterostructures also allows partial penetration of electric fields through the entire stack, thereby enabling gate tunability of the heterointerface.[160-162] The first demonstration along these lines was a tunneling field-effect transistor (TFET) that consists of a graphene/TMDC/graphene sandwich (Figure 7a),[160, 163] where the TMDC acts as a tunnel barrier (Figure 7b inset). In this structure, the tunneling probability can be controlled by tuning the Fermi level of the graphene by a gate voltage, thereby modulating the tunneling current between the two graphene layers by up to 6 orders of magnitude (Figure 7b).[160] This high current on/off ratio was confirmed independently by theory.[164] Heterostructure TFETs were subsequently fabricated on flexible substrates, where they exhibited high resilience to bending.[160] Likewise, a vertical heterojunction between a semimetal (graphene) and a semiconductor ($MoS_2$) creates a two-dimensional Schottky barrier at the interface that can be tuned by a gate voltage.[161] This concept was extended further to create a functional inverter by vertical stacking n-type and p-type layers separated by graphene layers as interconnects.[165]

The charge transport in graphene/TMDC/graphene structures and their variants occurs via a combination of tunneling (through the barrier) and thermionic (over the barrier) modes. The dominant mechanism depends on the thickness of the TMDC semiconductor, gate voltage, and temperature.[160, 163, 165] At lower TMDC thicknesses (1-5 nm), tunneling through the barrier dominates and thus a thicker TMDC (> 5nm) leads to lower currents in the 'off' state (negative gate voltages) and high on/off ratios.[160] For positive gate voltages and higher temperatures, the thermionic current over the barrier dominates charge transport. However, increased screening in thicker TMDCs leads to poor gate modulation of the top layer (further from the gate electrode) barrier thereby limiting on-current density.[165] This tradeoff between on/off ratio and on-current density can be partially circumvented using a metal contact at one end (instead of graphene) to pin the TMDC Fermi level.[165]

The absence of dangling bonds on TMDC basal planes presents additional opportunities to combine them with amorphous materials such as oxides, polymers, and organic small molecules. For example, ultrathin memory devices have been demonstrated using such combinations.[166-168] The earliest report of a semiconducting TMDC memory device involved using a ferroelectric polymer as the top dielectric and charge trapping layer.[168] Later demonstrations used graphene as both the contact and floating gate with $MoS_2$ as the channel and ALD-derived alumina as the tunnel barrier.[166] An alternative geometry employed h-BN as a tunnel barrier sandwiched between a $MoS_2$ FET and a graphene floating gate (Figure 7c,d).[167] These memory cells outperformed organic semiconductor-based memories in terms of retention time and endurance.[167] Bulk heterojunctions of chemically exfoliated $MoS_2$ and graphene oxide (GO) have also been used to fabricate memory devices albeit with much poorer performance compared to the planar heterostructures discussed above.[169, 170]

While TFETs, memory cells, and inverters have been realized using semiconducting TMDCs in layered heterostructures, the p-n junction is relatively unexplored. Note that p-n junctions based on bulk semiconducting TMDCs have been known for decades,[171, 172] but the lack of controlled doping strategies and the limited availability of ultrathin p-type semiconducting TMDCs has thus far prevented the realization of purely two-dimensional p-n heterojunctions. However, a p-n heterojunction can be fabricated by vertical stacking an n-type single-layer TMDC (*e.g.*, $MoS_2$, $WS_2$, or $MoSe_2$) with another p-type semiconductor such as semiconducting single-walled carbon nanotubes (s-SWCNTs)[173] or other conventional semiconductors.[174] For example, a gate-tunable p-n heterojunction diode has been achieved by vertical stacking n-type SL-$MoS_2$ and p-type semiconducting SWCNT thin films (Figure 7e).[173] The resulting p-n heterojunction diode not only shows gate-tunable rectification (Figure 7f inset) but also exhibits a unique '*anti-ambipolar*' transfer curve with two off states and a conductance maximum in between them (Figure 7f).[173] While charge transport mechanisms in all 2D materials such as graphene/TMDC heterostructures have been investigated to some extent as discussed above, the mechanisms in heterojunctions of mixed dimensionality components are not well understood. Further work will thus be required to understand issues such as band alignment, depletion regions, excess carrier separation, and interfacial charge transport mechanisms in these atomically thin p-n heterojunctions.

Gate tunability is not limited to the vertically stacked heterostructure geometry. In fact, each side of a lateral heterojunction can be independently tuned by individually addressable gates. As discussed above, the ultrathin nature of semiconducting TMDCs enables the subthreshold swing to be less than 80 mV/dec in top-gated SL-MoS$_2$ FETs at room temperature, thus making TMDCs attractive candidates for inter-band tunneling FETs. Initial theoretical work has proposed potential device geometries and indicated that a lower band gap semiconducting TMDC would be the most suitable candidate.[175, 176] Some efforts to create lateral p-n junctions from MoS$_2$ have been made, notably by ion gel gating an ambipolar device such that $V_{ds} > V_{gs}$[177] ($V_{ds}$ and $V_{gs}$ represent voltages applied between drain-source and gate-source contacts, respectively), and selective p-type doping of an MoS$_2$ flake by exposure to various plasmas.[178, 179] In-plane p-n heterojunctions from two disparate semiconducting TMDCs are also an enticing possibility considering that ternary two-dimensional alloys of Mo$_x$W$_{1-x}$S$_2$ and MoS$_{2(1-x)}$Se$_{2x}$ have already been synthesized.[180-183] It is, however, expected that precise control over growth and lattice matching will present issues such as chalcogen intermixing and control over dopant density in selective areas, and thus in-plane engineering of two-dimensional heterostructures and p-n junctions may be more constrained than their vertical heterojunction counterparts. Since the band structures of semiconducting TMDCs undergo dramatic changes with thicknesses in the 1-6 layer limit,[44] thickness-dependent lateral junctions are another possibility. For example, one such lateral junction between monolayer and bilayer graphene exhibited a photothermoelectric response.[184] Furthermore, some approaches to create thickness-dependent lateral TMDC junctions are already in place,[185-187] although no working devices have yet been demonstrated.

## **Optoelectronics**

With direct band gaps in the visible portion of the electromagnetic spectrum, large exciton binding energies, and strong photoluminescence, monolayer semiconducting TMDCs are promising for optoelectronic applications. Several classes of optoelectronic devices have been demonstrated from ultrathin semiconducting TMDCs including photodetectors, photovoltaics, and light-emitting devices. This section will review these applications in detail.

### **Photodetectors**

When photons of energy greater than the band gap are incident on a semiconductor, they create bound electron-hole pairs (excitons) or free carriers depending on the exciton binding energy in the semiconductor. Bound excitons separated by an applied or built-in electric field generate a photocurrent. Two major categories of semiconductor-based photodetectors exist, namely phototransistors and photodiodes.

Initial studies on ultrathin $MoS_2$-based phototransistors measured photocurrent under global illumination[188, 189] with photons possessing energies greater than the 1.9 eV band gap (Figure 8a).[189] Phototransistors have also been recently extended to chemical vapor deposition grown few-layer $WS_2$.[190] Wavelength-dependent studies later suggested that the photocurrent roughly follows the absorption spectrum, which led to speculation of interband absorption and carrier separation as the dominant mechanism of photocurrent generation.[188, 191] In this context, scanning photocurrent microscopy has emerged as a powerful tool to elucidate the underlying photocurrent generation mechanisms in nanoscale devices.[192, 193] An initial report of scanning photocurrent measurements on unbiased SL-MoS$_2$ devices claimed that the photothermoelectric effect is the dominant mechanism for photocurrent generated near the contacts.[116] Subsequent studies on the bias dependence of spatially resolved photocurrents in MoS$_2$ FETs suggested interband excitation as the dominant photocurrent mechanism under bias (Figure 8b,c), although the relative contributions from the two mechanisms at zero bias may vary between devices.[194] At large drain biases (*i.e.*, saturation regime), the electric field is large enough to separate photoexcited electron-hole pairs across the channel, thereby leading to a larger net photocurrent in the channel (Figure 8d).[195] Most phototransistors operating under bias achieve larger than unity responsivity (often R > $10^3$ A/W) *via* internal amplification of the photocurrent. For example, the graphene-MoS$_2$ vertical heterostructure channel led to phototransistors with R > $10^7$ A/W (Figure 8d),[196, 197] similar in concept to quantum dot sensitized graphene photodetectors.[198] Despite such high responsivities, phototransistors suffer from relatively long response times compared to photodiodes.[199]

A photodetector can have short response times if the photogenerated carriers are separated and collected within a short distance by a built-in electric field. For ultrafast photodetectors, built-in fields are achieved *via* a variety of junctions including semiconductor-metal Schottky junctions, homojunctions between differently doped regions in a semiconductor, and heterojunctions between two different semiconductors. Although Schottky junctions are the main source of the low bias photoresponse in most reports of MoS$_2$ phototransistors, the concept of a metal-semiconductor-metal (M-S-M) junction has been systematically exploited only recently.[200, 201] This approach enables a faster response time as compared to conventional FETs primarily due to the reduced channel length between the two M-S junctions. In particular, intrinsic speed limits of a M-S-M photodetector can be achieved if the channel length is reduced to less than twice the depletion width. Such photodetectors comprising vertical stacks of graphene/TMDC/graphene have been demonstrated with response speeds down to 50 μs.[202] Nevertheless, a p-n heterojunction photodiode presents opportunities for even faster photodetection. Heterojunctions of p-type Si (crystalline[203] and amorphous[204]) with n-type $MoS_2$ have now been realized. In addition, the aforementioned gate-tunable p-n heterojunction diode based on p-type semiconducting SWCNTs and n-type SL-$MoS_2$ exhibits a fast photoresponse (rise time < 15 μs, limited by instrumentation) (Figure 8e) with peak responsivity of ~0.1 A/W.[173] Van der Waals p-n heterojunctions between semiconducting TMDCs and other direct band gap materials are expected to show similar results.

**Solar cells and light-emitting devices**

The photovoltaic cell is one of the most widespread applications of a p-n junction. As per the Shockley-Quessier limit, a high mobility semiconductor with a direct band gap near 1.3 eV is desired for high efficiency single junction photovoltaics. Single-layer semiconducting TMDCs are well-positioned here in terms of both direct band gap and mobility values (Figure 9), making them promising candidates for photovoltaic applications. Among several recent theoretical studies, one calculation on a Schottky junction solar cell consisting of a graphene-$MoS_2$ stack suggests a maximum power conversion efficiency (PCE) of ~1% while that of a type-II heterojunction between $WS_2$/$MoS_2$ is 1.5%.[205] Experimentally, an asymmetric M-S-M Schottky junction on a few-layer $MoS_2$ flake resulted in working photovoltaic devices with ~1 % PCE,[206] while split-gated p-n junctions on monolayer WSe$_2$ achieved a PCE of 0.5%.[207] Recently, vertically stacked graphene/TMDC/graphene structures have also shown promise with external quantum efficiency values as high as 55% at a single wavelength.[202] Although the above results for ultrathin solar cells based on semiconducting TMDCs are encouraging, thickness-limited absorption poses a challenge for high efficiency. To overcome this issue, light trapping techniques such as plasmon-enhanced absorption,[208-210] strain engineering,[211] and/or vertical stacks of atomically thin cells are being pursued.

The light-emitting diode (LED) is another ubiquitous application of the p-n junction. As direct gap materials, monolayer semiconducting TMDCs are a promising option for ultrathin, efficient, and flexible LEDs. However, the lack of controlled doping strategies has thus far hindered LEDs fabricated from purely monolayer TMDCs. Nevertheless, Schottky junction based light-emitting devices have been realized albeit with low quantum efficiencies ($\sim10^{-5}$).[212] Based on the heterojunction photodetection mechanism mentioned above, excitonic electroluminescence has also been observed from SL-MoS$_2$ and crystalline p-type Si heterojunctions.[203] Homojunctions formed by electrostatic doping of nearly intrinsic monolayer WSe$_2$ have also resulted in atomically thin LEDs with maximum efficiencies of ~0.2%.[207, 213, 214] With commercial organic LEDs approaching 20% emission efficiencies, improvements in doping, surface engineering, encapsulation, and device design are needed for semiconducting TMDCs to be competitive light emitters.

**Sensors**

Electronic sensors exploit changes in the electrical properties of constituent materials upon interaction with an analyte. When a bulk material in a FET channel interacts with an analyte, its resistance or the device capacitance typically undergo relatively small changes since the interaction is limited to the surface. However, in atomically thin nanomaterials, the surface-to-volume ratio is close to unity and thus the electrical properties are significantly perturbed by even submonolayer analyte adsorption (Figure 3b). This strategy has been heavily exploited in carbon nanomaterials[3] and should also apply to ultrathin semiconducting TMDCs.

Initial studies on TMDC-based gas sensing employed $MoS_2$ FETs for the detection of nitrogen monoxide (NO). The resistance of single-layer and few-layer $MoS_2$ FETs was found to be sensitive to the presence of NO down to 0.8 ppm in an $N_2$ environment.[215] A subsequent study focused on elucidating the effect of gate voltage (*i.e.*, carrier concentration), layer thickness, and analyte type (reducing or oxidizing) on the sensing properties. It was found that multi-layer $MoS_2$ (5 layers) was the most sensitive to all of the analytes under consideration (*i.e.*, $NO_2$, $NH_3$, and $H_2O$). In addition, a positive gate bias (*i.e.*, accumulation) has the opposite effect on the sensitivity of reducing ($NH_3$) *versus* oxidizing ($NO_2$) analytes.[216] The authors concluded that charge transfer to the $MoS_2$ channel underlies the sensing mechanism, although some ambiguity remains regarding the effect of contact doping as observed in the case of $WSe_2$ FETs.[125] Another detailed study using small organic molecule analytes observed that the sensitivity of the SL-$MoS_2$ FET channel (Figure 10a) was highest for the most reducing and polar analytes while the channel was insensitive to oxidizing analytes. Such selectivity was not observed for the comparative case of carbon nanotube network sensors (Figure 10b).[217] While all of the above studies were performed on mechanically exfoliated flakes, attempts at scaling up $MoS_2$ chemical sensors have been recently made using chemical vapor deposition films[218] and solution-phase exfoliated flakes.[219-221]

In addition to sensing based on changes in resistance, monolayer semiconducting TMDCs also provide an alternative avenue for sensing based on light emission. In particular, the physisorption and resulting charge transfer upon analyte binding leads to changes in the PL/EL emission efficiencies depending on the polarity of the semiconductor. This strategy has been demonstrated using electron-withdrawing analytes (*e.g.*, $O_2$ and $H_2O$), which enhance the PL efficiencies of n-type $MoS_2$ and $WS_2$ (Figure 10c) while reducing PL efficiencies for p-type $WSe_2$.[222] The strain-dependent shift in band structure and exciton binding energies of monolayer semiconducting TMDCs have also led to attempts at exploiting this phenomenon for strain sensing (Figure 10d).[223]

Although the field of semiconducting TMDC-based electronic devices has evolved rapidly, studies on sensor development have been relatively limited. Specifically, the issues of selectivity and reset-ability require significantly more attention. Since the performance degradation of TMDC devices due to adsorbate interaction is reversible by subjecting the devices to vacuum conditions, the reset-ability of TMDC based sensors is likely to be superior to organic semiconductors where this degradation is irreversible. Following previous work on carbon nanomaterials,[3] chemical selectivity can likely be achieved through chemical functionalization methods. Of particular interest will be functionalization strategies that are chemically robust and yet do not significantly perturb the electronic properties of the underlying TMDCs.

**Conclusions and outlook**

Significant progress has been made in many application areas of semiconducting TMDCs. The bulk of this effort has focused on FETs, where the underlying charge transport and scattering mechanisms have been identified and the effects of dielectrics, environment, and contacts have been delineated. This fundamental knowledge has begun to translate into rudimentary functional electronic circuits, although integration issues such as complementary doping and threshold voltage control require more effort. Similarly, access to high-quality, large-area substrates is emerging with the advent of chemical vapor deposition grown material, although improvements are needed for high-performance circuit applications. For optoelectronics, methods for enhancing light absorption and improving fluorescence and electroluminescence quantum yields in ultrathin materials will be needed if they are to be competitive with conventional bulk semiconductors. Finally, for sensing applications, chemical functionalization methods are desired that impart high chemical selectivity and robustness without disrupting the superlative electronic properties of the TMDC semiconductor.

With regard to the commercialization potential for semiconducting TMDCs, it appears that unconventional formats that require flexibility, stretchability, and/or transparency are most promising. Since even a monolayer (thickness < 1 nm) semiconducting TMDC can give a comparable electrical performance to a 10 nm thick organic or amorphous oxide semiconductor, ultrathin TMDCs are particularly well-suited for transparent and flexible electronics. However, with high temperature (> 400 ºC) chemical vapor deposition as the only known method to obtain electronic grade TMDCs over large areas, direct growth or transfer onto flexible/transparent plastic substrates will pose challenges. Similarly, solution-based methods for preparing and depositing TMDC materials require substantial improvement, especially for high performance electronic and optoelectronic applications.

While many applications of semiconducting TMDCs are virtually identical to other electronic materials, the atomically thin nature of TMDCs presents unique opportunities. In particular, the gate tunability of heterostructure devices enables fundamentally different charge transport phenomena including the anti-ambipolarity that has been observed in SWCNT/SL-MoS$_2$ p-n heterojunctions. In such heterostructure devices, the spectral photoresponse can also be tuned with an external gate bias, enabling dynamic tailoring of device characteristics. Since this gate tunability does not exist in bulk semiconductors, devices that exploit it are unlikely to face competition from conventional materials and thus present unimpeded access to new markets. By focusing directly on such unique opportunities, the technological impact of semiconducting TMDCs can likely be maximized.

**Conflict of interest:** The authors declare no competing financial interests.

ACKNOWLEDGMENT: This work was supported by the Materials Research Science and Engineering Center (MRSEC) of Northwestern University (National Science Foundation Grant DMR-1121262) and the Office of Naval Research Multidisciplinary University Research Initiative Program (Grant N00014-11-1-0690).

**FIGURES**

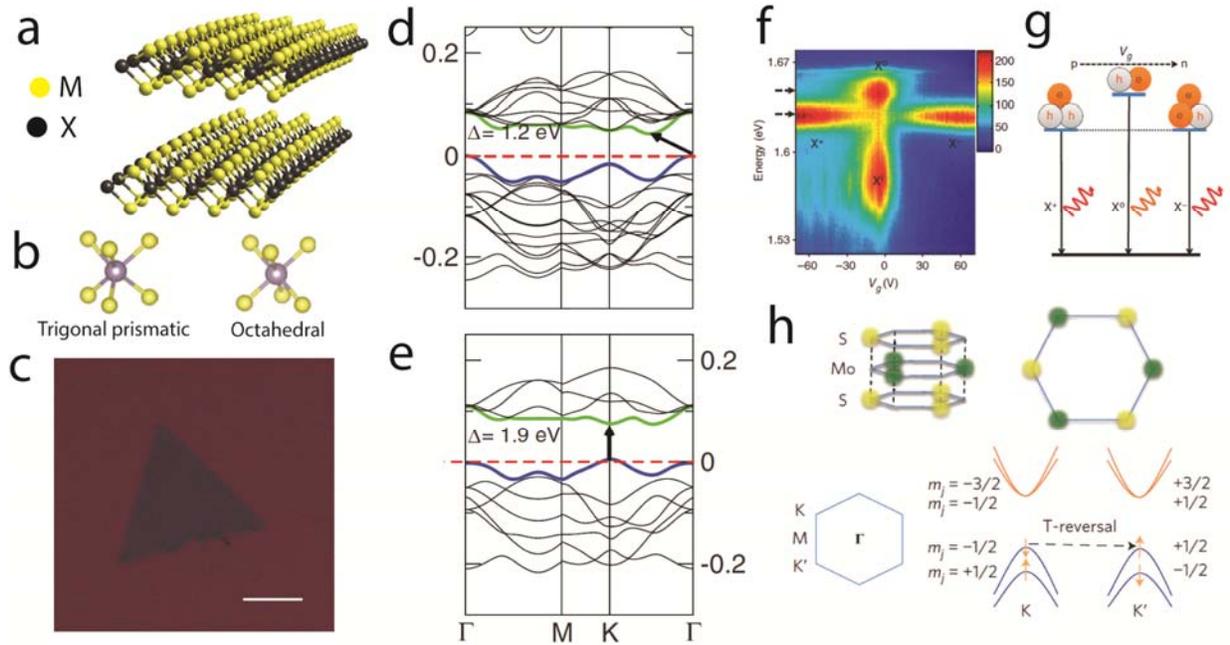

**Figure 1.** Physical properties of layered semiconducting transition metal dichalcogenides (TMDCs). (a) Chemical structure of two layers of a TMDC where M is a transition element and X is a chalcogen. (b) Two polytypes of single-layer TMDCs: trigonal prismatic (1H) and octahedral (1T). Adapted from ref.[8] © 2013, Macmillan Publishers Ltd. (c) Optical micrograph of single-layer $MoS_2$ on a 300 nm $SiO_2$/Si substrate (scale bar = 20 μm). (d) Band structure of bulk $MoS_2$ and (e) single-layer $MoS_2$ as calculated from density functional theory. Adapted from ref.[46] © 2011, American Physical Society. (f) Photoluminescence energy of single-layer $MoSe_2$ plotted as a function of gate voltage. At zero gate voltage, there are neutral ($X^0$) and impurity-trapped excitons ($X^I$). With large electron and hole doping, charged excitons (trions) dominate the spectrum. (g) Schematic of a gate-dependent transition of a positive trion ($X^+$) to a neutral exciton and then to a negative trion ($X^-$). Adapted from ref.[56] © 2013, Macmillan Publishers Ltd. (h) Top: Honeycomb lattice of single-layer $MoS_2$ where alternating corners are occupied by one

Mo and two S atoms, resulting in broken spatial inversion symmetry. Bottom: The conduction band minima and valence band maxima are shown with corresponding z-components of their total angular momentum. The valence bands are split due to strong spin-orbit coupling. Adapted from ref.[64] © 2012, Macmillan Publishers Ltd.

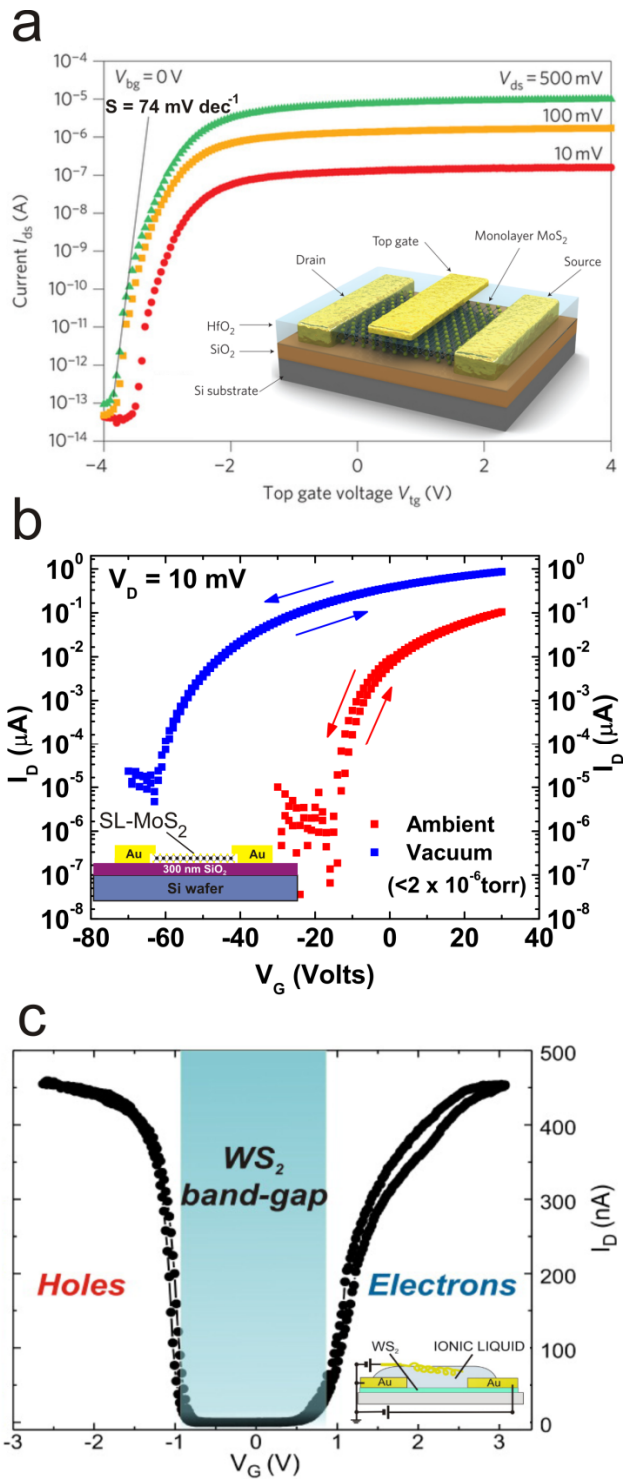

**Figure 2**. Gate dielectric choices for semiconducting TMDC FETs. (a) Semi-log transfer characteristics of a top-gated SL-MoS$_2$ FET with a 30 nm thick hafnia dielectric. The low sub-

threshold swing, low operating voltage, and high on/off ratio result from enhanced gate coupling. The inset shows a device schematic. Adapted from ref.[16] © 2011, Macmillan Publishers Ltd. (b) Semi-log transfer curves of an unencapsulated bottom gated SL-MoS$_2$ FET measured under ambient (red) and vacuum (2 x 10$^{-6}$ Torr) (blue), showing that atmospheric adsorbates have deleterious effects on device performance. The inset shows a device schematic. Adapted from ref.[82] © 2013, American Institute of Physics. (c) Ambipolar transfer characteristic of an ion-gel gated few-layer WS$_2$ FET. The band gap of the sample can be estimated from the width of the off-state. The inset shows the measurement schematic. Adapted from ref.[94] © 2012, American Chemical Society.

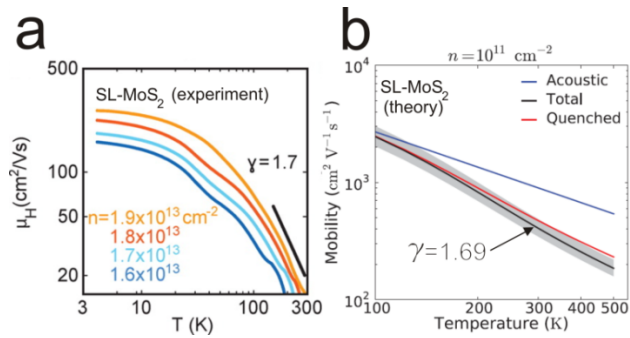

**Figure 3.** Temperature dependence of carrier mobility. (a) Hall mobility values in single-layer MoS$_2$ at different carrier densities as a function of temperature. The temperature exponent γ obtained from the experimental data agrees well with the theoretical estimate in (b). Adapted from ref.[102] © 2013, American Chemical Society. (b) Calculated carrier mobility as a function of temperature for single-layer MoS$_2$. The black curve represents the temperature dependence of mobility considering charge scattering by both acoustic and optical phonons. The red curve shows mobility when out-of-plane optical phonons are quenched, while the blue curve considers only acoustic phonons. Adapted from ref.[105] © 2012, American Physical Society.

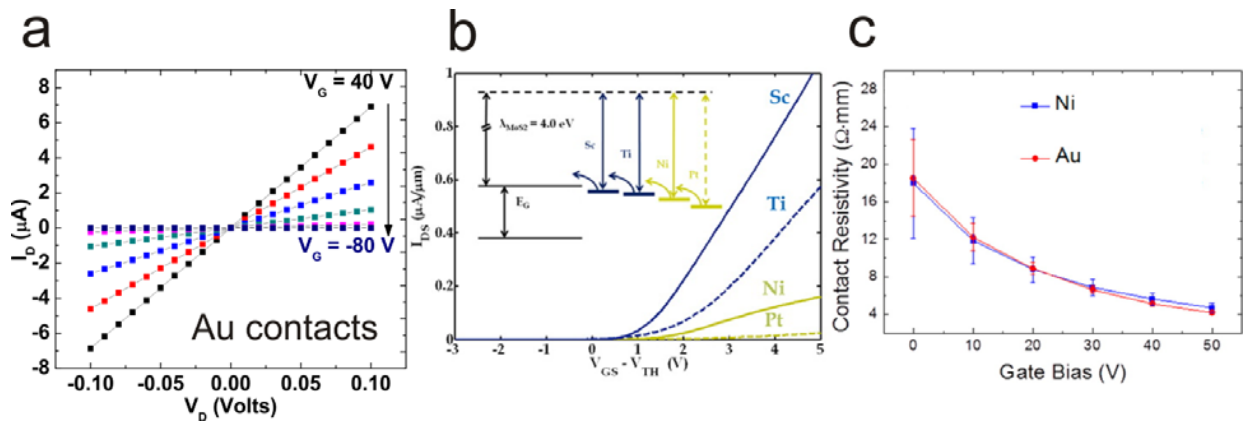

**Figure 4.** Metal contacts. (a) Linear $I_d$-$V_{ds}$ output curves in an unencapsulated, bottom-gated single-layer MoS$_2$ FET with Au contacts. Adapted from ref.[82] © 2013, American Institute of Physics. (b) Increase in drain current of a few-layer MoS$_2$ FET with decreasing work function of the contact metals. Adapted from ref.[99] © 2012, American Chemical Society. (c) Gate voltage dependence of the contact resistance in a few-layer MoS$_2$ FET indicative of a Schottky barrier at the contacts. Adapted from ref.[129] © 2012, American Chemical Society.

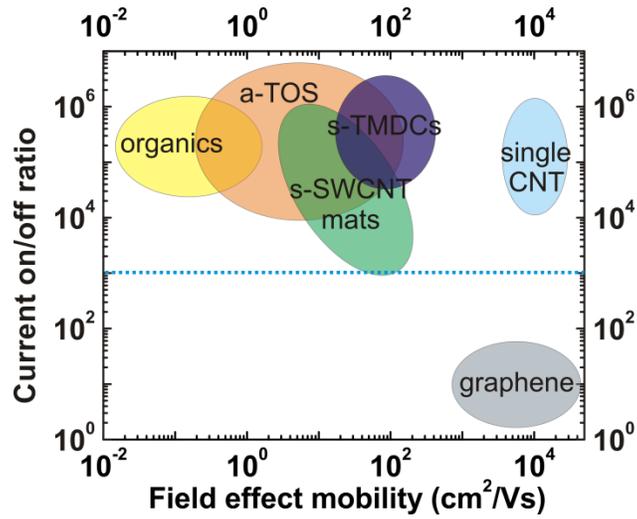

**Figure 5.** Comparison of field-effect mobility and on/off ratios of semiconductors for unconventional electronics. High on/off ratios (>10$^3$) are essential for digital electronics applications while maintaining the highest mobility possible. Semiconducting TMDCs (s-TMDCs) lie on a favorable position on this plot compared to other semiconductors that are currently under consideration for large-area flexible electronics.

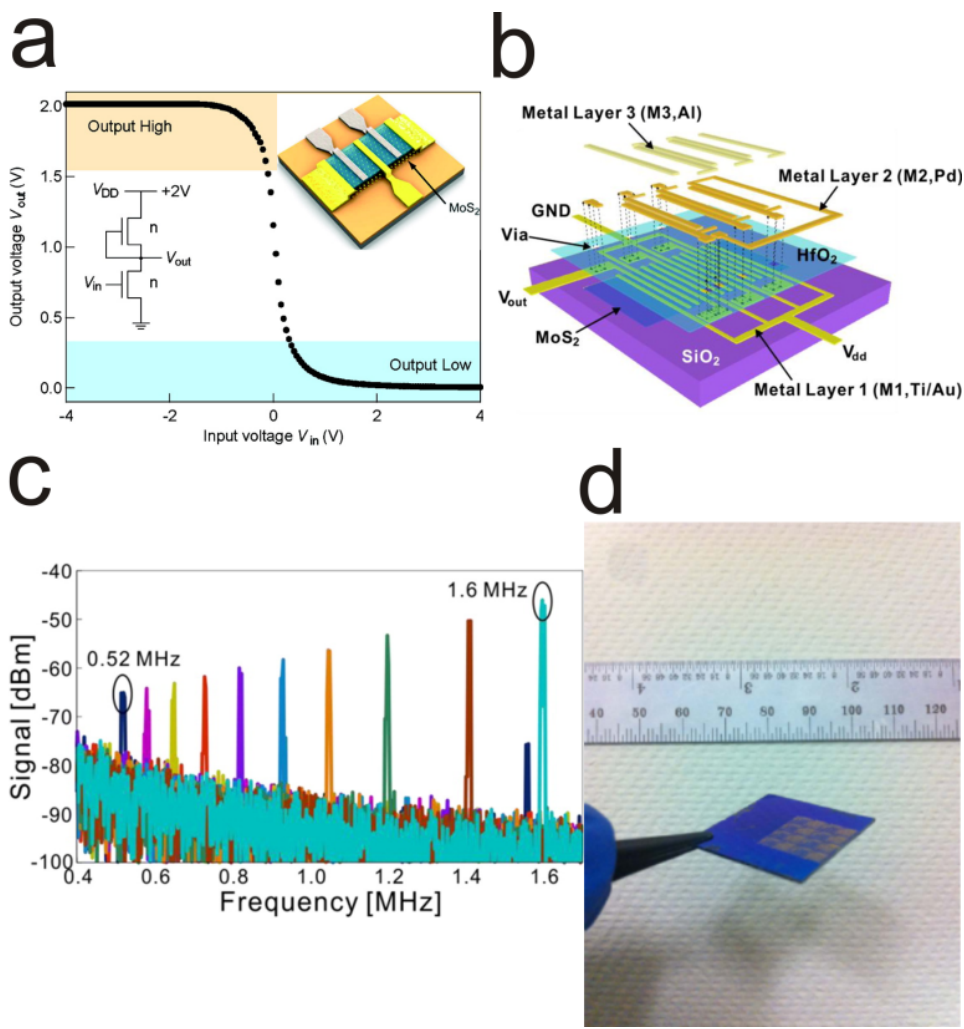

**Figure 6.** Integrated circuits from ultrathin semiconducting TMDCs. (a) Output voltage as a function of input voltage of an inverter based on two single-layer MoS$_2$ FETs. The inset shows the circuit diagram (left) and device schematic (right). Adapted from ref.[141] © 2011, American Chemical Society. (b) Schematic illustration of a ring oscillator circuit fabricated on ultrathin MoS$_2$. (c) Power spectrum of the ring oscillator output signal as the supply voltage ($V_{dd}$) is increased from left to right. Adapted from ref.[142] © 2012, American Chemical Society. (d) Demonstration of electronic devices on chemical vapor deposition grown single-layer MoS$_2$ over large areas. Adapted from ref.[143] © 2012, Institute of Electrical and Electronics Engineers.

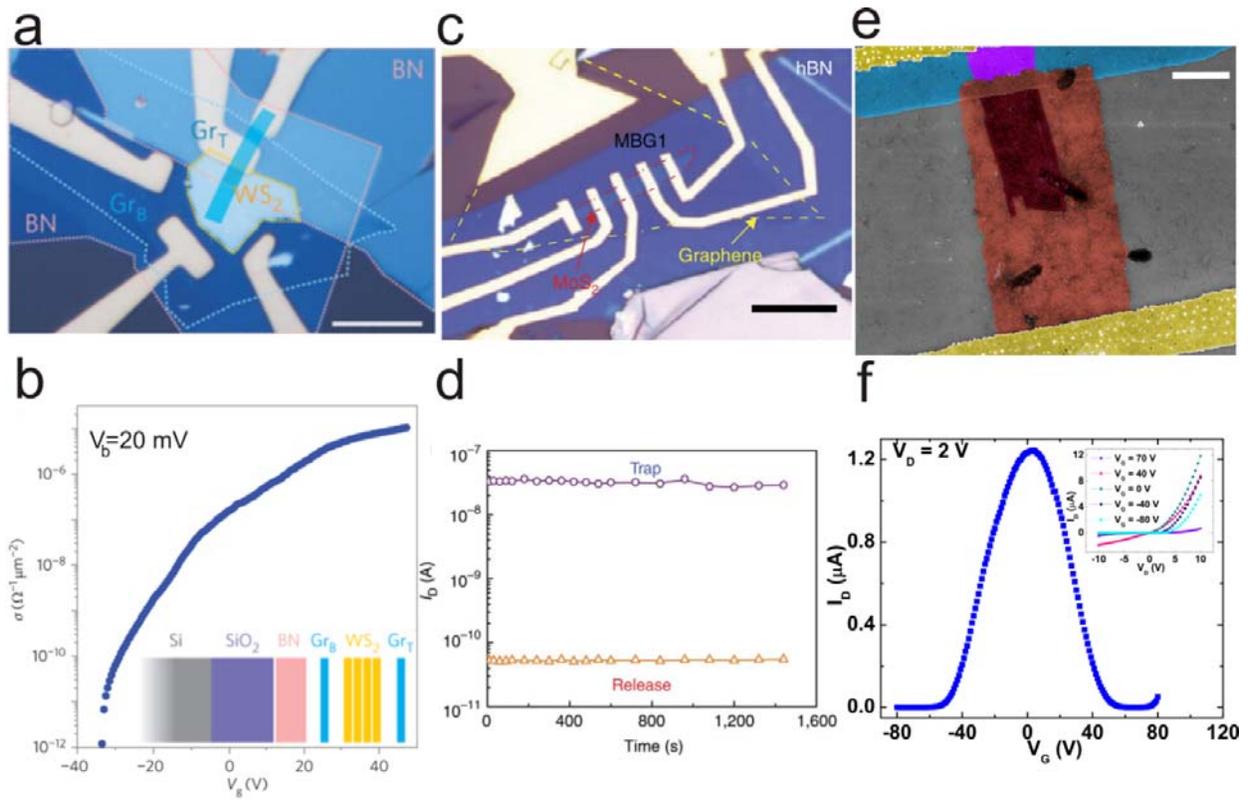

**Figure 7.** Van der Waals heterostructure devices. (a) Optical micrograph of a tunneling field-effect transistor (TFET) composed of a graphene/WS$_2$/graphene sandwich encapsulated between h-BN layers on the top and bottom. Gr$_b$ and Gr$_t$ stand for bottom and top graphene, respectively. (b) Semi-log transfer curve of the same device showing on/off ratios > 10$^6$. The inset shows a schematic diagram of the device cross-section. Adapted from ref.[160] © 2013, Macmillan Publishers Ltd. (c) Optical micrograph of a MoS$_2$/h-BN/graphene memory cell with the layer boundaries appropriately colored. (d) Temporal current stability through the MoS$_2$ channel of the same device after applying the write (trap) and erase (release) pulses. Adapted from ref.[167] © 2013, Macmillan Publishers Ltd. (e) False-colored scanning electron micrograph of a s-SWCNT/SL-MoS$_2$ p-n heterojunction. The dark red region is the MoS$_2$ flake, which is colored

pink as it goes under the atomic layer deposition grown alumina (blue). The pink rectangle is a patterned film of randomly oriented semiconducting SWCNTs. (f) Anti-ambipolar transfer curve of the same device with two off states and a current peak between them. The inset shows the gate dependence of the output curves of the diode ranging from an insulating to a highly rectifying state. Adapted from ref.[173] © 2013, National Academy of Sciences, U.S.A.

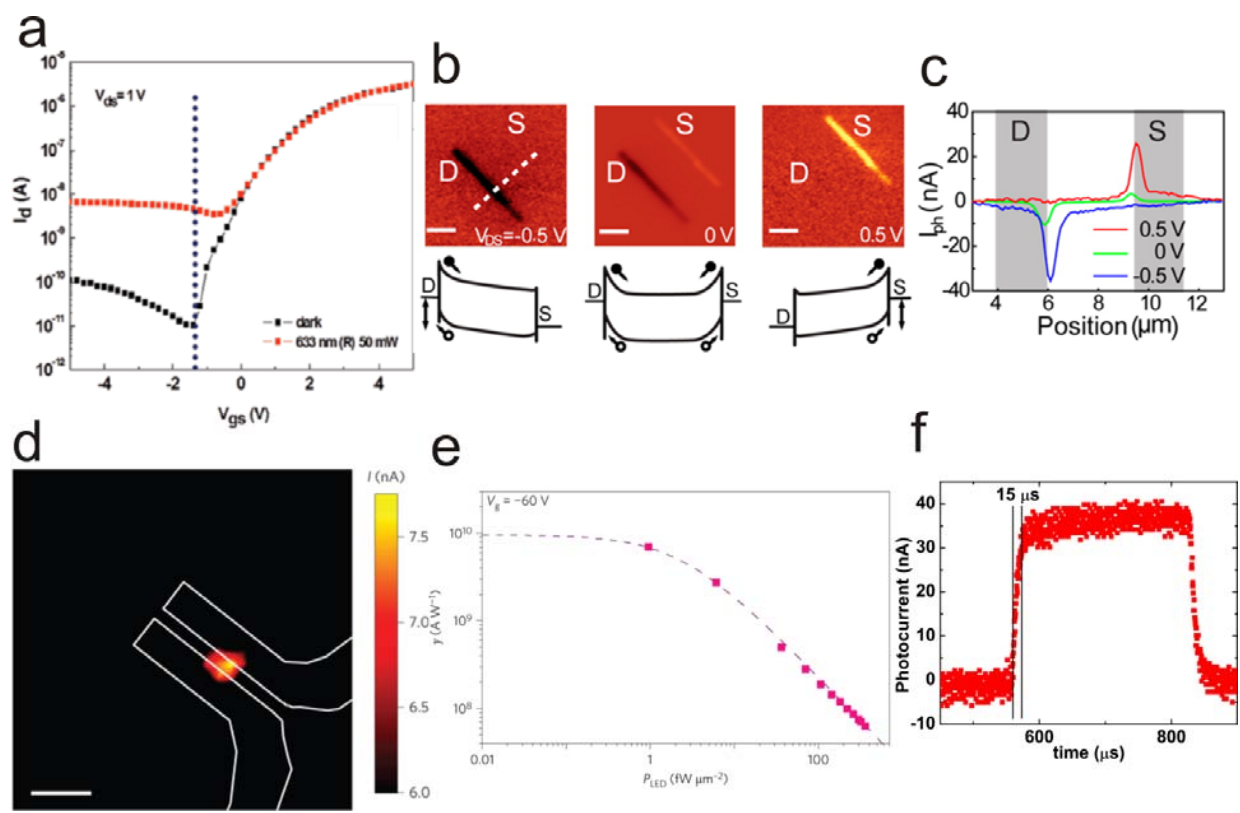

**Figure 8.** Photodetectors from s-TMDCs. (a) Semi-log transfer curves of a few-layer $MoS_2$ photo-FET in dark (black) and under illumination (red). Photocurrent is seen in depletion mode. The inset shows the device schematic Adapted from ref.[189] © 2012, Wiley-VCH. (b) Spatial photocurrent maps of a few-layer $MoS_2$ FET under varying $V_{ds}$. The photocurrent magnitude at the source (drain) increases (decreases) under forward bias and *vice versa* for reverse bias with corresponding schematic band diagrams below. (c) Line profiles of the photocurrents in (b) along the dashed white line. Adapted from ref.[194] © 2013, American Chemical Society. (d) Spatial photocurrent map of a single-layer $MoS_2$ FET under bias. Drift field induced photocurrent generation from the entire channel can be seen. The inset shows the schematic of the device with laser beam. Adapted from ref.[195] © 2013, Macmillan Publishers Ltd. (e) Photoresponsivity

*versus* incident power of a graphene/MoS$_2$ bilayer photo-FET. The inset shows the device schematic. Adapted from ref.[196] © 2013, Macmillan Publishers Ltd. (f) Fast photoresponse from a s-SWCNT/SL-MoS$_2$ p-n heterojunction diode. An instrument-limited rise time of 15 μs can be seen. Adapted from ref.[173] © 2013, National Academy of Sciences, U.S.A.

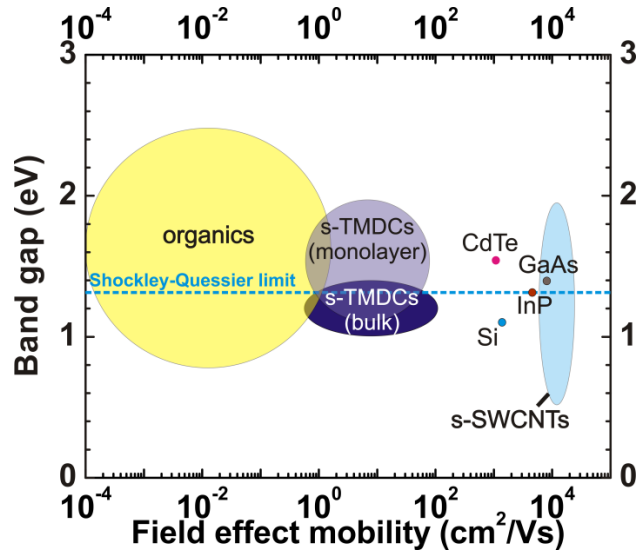

**Figure 9.** Band gap versus field-effect mobility for important semiconductors used in current generation photovoltaic technologies. A band gap near the Shockley-Quessier limit (~1.3 eV) and high mobilities is expected to lead to high efficiency photovoltaic devices. Semiconducting TMDCs are favorably placed in comparison to organic semiconductors for flexible photovoltaic cells.

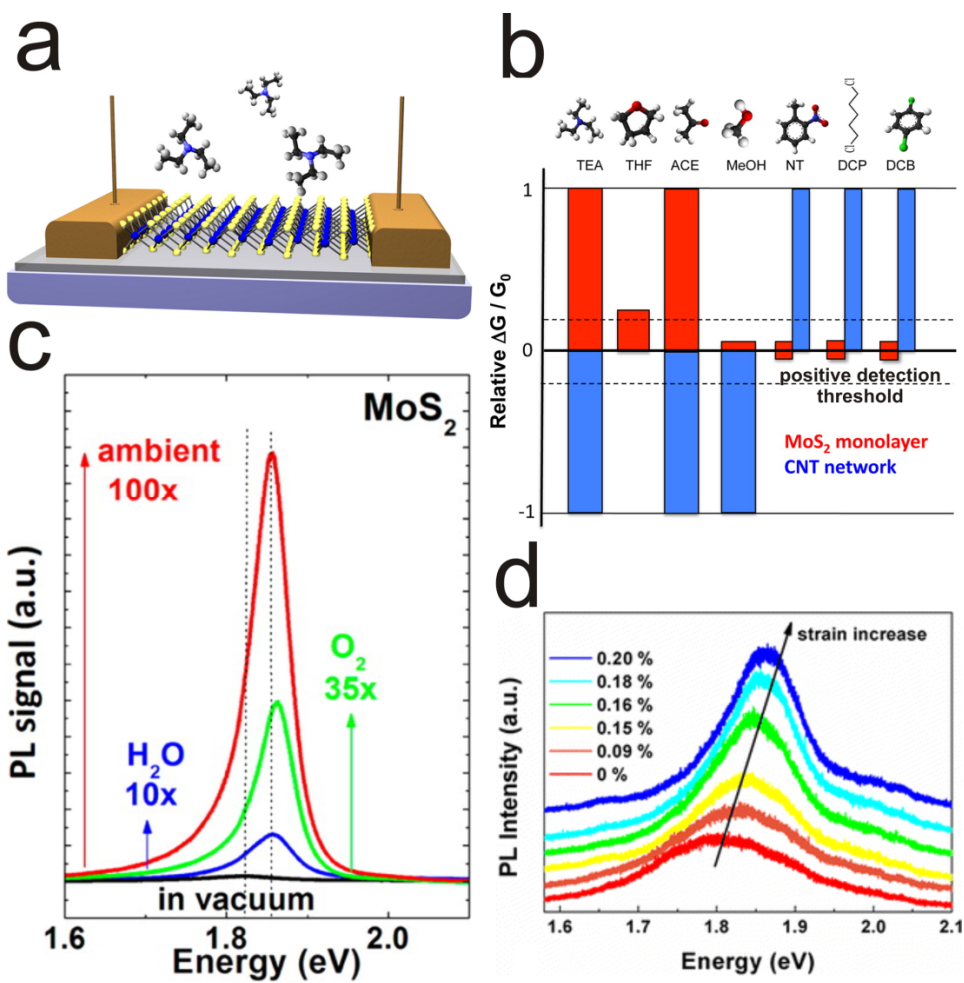

**Figure 10.** Sensors from semiconducting TMDCs. (a) Schematic of a single-layer $MoS_2$ FET sensing small organic analytes. (b) Sensitivity of a single-layer $MoS_2$ FET (red) compared with a random network SWCNT FET. The higher selectivity of single-layer $MoS_2$ over SWCNTs can be seen. Adapted from ref.[217] © 2013, American Chemical Society. (c) Photoluminescence (PL) efficiency of single-layer $MoS_2$ in the presence of various gaseous adsorbates. Adapted from ref.[222] © 2013, American Chemical Society. (d) PL spectra from a single-layer $MoS_2$ FET under varying amount of strain. Adapted from ref.[223] © 2011, American Chemical Society.
40

188. Yin, Z.; Li, H.; Li, H.; Jiang, L.; Shi, Y.; Sun, Y.; Lu, G.; Zhang, Q.; Chen, X.; Zhang, H. Single-Layer MoS$_2$ Phototransistors. *ACS Nano* 2012, 6, 74-80.
189. Choi, W.; Cho, M. Y.; Konar, A.; Lee, J. H.; Cha, G. B.; Hong, S. C.; Kim, S.; Kim, J.; Jena, D.; Joo, J., *et al.* High-Detectivity Multilayer MoS$_2$ Phototransistors with Spectral Response from Ultraviolet to Infrared. *Adv. Mater.* 2012, 24, 5832-5836.
190. Perea-López, N.; Elías, A. L.; Berkdemir, A.; Castro-Beltran, A.; Gutiérrez, H. R.; Feng, S.; Lv, R.; Hayashi, T.; López-Urías, F.; Ghosh, S., *et al.* Photosensor Device Based on Few-Layered WS$_2$ Films. *Adv. Funct. Mater.* 2013, 23, 5511–5517.
191. Lee, H. S.; Min, S.-W.; Chang, Y.-G.; Park, M. K.; Nam, T.; Kim, H.; Kim, J. H.; Ryu, S.; Im, S. MoS$_2$ Nanosheet Phototransistors with Thickness-Modulated Optical Energy Gap. *Nano Lett.* 2012, 12, 3695-3700.
192. Allen, J. E.; Hemesath, E. R.; Lauhon, L. J. Scanning Photocurrent Microscopy Analysis of Si Nanowire Field-Effect Transistors Fabricated by Surface Etching of the Channel. *Nano Lett.* 2009, 9, 1903-1908.
193. Gu, Y.; Kwak, E. S.; Lensch, J. L.; Allen, J. E.; Odom, T. W.; Lauhon, L. J. Near-field scanning photocurrent microscopy of a nanowire photodetector. *Appl. Phys. Lett.* 2005, 87, 043111.
194. Wu, C.-C.; Jariwala, D.; Sangwan, V. K.; Marks, T. J.; Hersam, M. C.; Lauhon, L. J. Elucidating the Photoresponse of Ultrathin MoS$_2$ Field-Effect Transistors by Scanning Photocurrent Microscopy. *J. Phys. Chem. Lett.* 2013, 4, 2508-2513.
195. Lopez-Sanchez, O.; Lembke, D.; Kayci, M.; Radenovic, A.; Kis, A. Ultrasensitive Photodetectors Based on Monolayer MoS$_2$. *Nat. Nanotechnol.* 2013, 8, 497–501.
196. Roy, K.; Padmanabhan, M.; Goswami, S.; Sai, T. P.; Ramalingam, G.; Raghavan, S.; Ghosh, A. Graphene-MoS$_2$ Hybrid Structures for Multifunctional Photoresponsive Memory Devices. *Nat. Nanotechnol.* 2013, 8, 826–830.
197. Zhang, W.; Chuu, C.-P.; Huang, J.-K.; Chen, C.-H.; Tsai, M.-L.; Chang, Y.-H.; Liang, C.-T.; He Jr, -. H.; Chou, M.-Y.; Li, L.-J. Ultrahigh-Gain Photodetectors Based on Atomically Thin Graphene-MoS$_2$ Heterostructures *Sci. Rep.* 2013, 4, 3826.
198. Konstantatos, G.; Badioli, M.; Gaudreau, L.; Osmond, J.; Bernechea, M.; de Arquer, F. P. G.; Gatti, F.; Koppens, F. H. Hybrid graphene-quantum dot phototransistors with ultrahigh gain. *Nat. Nanotechnol.* 2012, 7, 363-368.
199. Sze, S. M.; Ng, K. K. *Physics of Semiconductor Devices*. 3$^{rd}$ ed.; Wiley: 2007.
200. Tsai, D.-S.; Lien, D.-H.; Tsai, M.-L.; Su, S.-H.; Chen, K.-M.; Ke Jr, -. J.; Yu, Y.-C.; Li, L.-J.; He Jr, -. H. Trilayered MoS$_2$ Metal-Semiconductor-Metal Photodetectors: Photogain and Radiation Resistance. *IEEE J. Sel. Top. Quant.* 2014, 20, 3800206.
201. Tsai, D.-S.; Liu, K.-K.; Lien, D.-H.; Tsai, M.-L.; Kang, C.-F.; Lin, C.-A.; Li, L.-J.; He, J.-H. Few Layer MoS$_2$ with Broadband High Photogain and Fast Optical Switching for Use in Harsh Environments. *ACS Nano* 2013, 7, 3905–3911.
202. Yu, W. J.; Liu, Y.; Zhou, H.; Yin, A.; Li, Z.; Huang, Y.; Duan, X. Highly Efficient Gate-Tunable Photocurrent Generation in Vertical Heterostructures of Layered Materials. *Nat. Nanotechnol.* 2013, 8, 952–958.
203. Ye, Y.; Ye, Z.; Gharghi, M.; Zhu, H.; Zhao, M.; Yin, X.; Zhang, X. Exciton-Related Electroluminescence from Monolayer MoS$_2$. *arXiv:1305.4235* 2013.
204. Esmaeili-Rad, M. R.; Salahuddin, S. High Performance Molybdenum Disulfide Amorphous Silicon Heterojunction Photodetector. *Sci. Rep.* 2013, 3, 2345.
52

**TOC Image:**

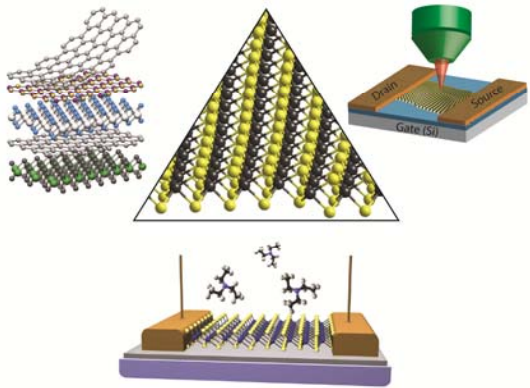

Emerging Device Applications of Semiconducting Two-Dimensional Transition Metal Dichalcogenides